\documentclass[aps,prd,12pt,amssymb]{revtex4}

\begin{document}

\baselineskip=0.60cm

\newcommand{\ini}{\begin{equation}}
\newcommand{\fin}{\end{equation}}
\newcommand{\inir}{\begin{eqnarray}}
\newcommand{\finr}{\end{eqnarray}}
\newcommand{\inif}{\begin{figure}}
\newcommand{\finf}{\end{figure}}
\newcommand{\bc}{\begin{center}}
\newcommand{\ec}{\end{center}}

\def\ol{\overline}
\def\pa{\partial}
\def\ra{\rightarrow}
\def\ts{\times}
\def\df{\dotfill}
\def\bs{\backslash}
\def\dg{\dagger}
\def\la{\lambda}

$~$

\hfill DSF-05/2005

\vspace{1 cm}

\title{LEPTON MIXING AND SEESAW MECHANISM}

\author{D. Falcone}

\affiliation{Dipartimento di Scienze Fisiche,
Universit\`a di Napoli, Via Cintia, Napoli, Italy}

\begin{abstract}
\vspace{1cm}
\noindent
In the context of a typical model for fermion mass matrices, possibly 
based on the horizontal U(2) symmetry, we explore the effect of the type 
II seesaw mechanism on lepton mixings. We find that the combined 
contribution of type I and type II terms is able to explain the 
large but not maximal 1-2 mixing and the near maximal 2-3 mixing, while 
the 1-3 mixing angle is predicted to be small.
\end{abstract}

\maketitle

\newpage

\section{Introduction}

By now there is convincing evidence for neutrino oscillation, which imply
neutrino mass and lepton mixing. Thus, the framework seems analogous to
that of quark mass and mixing. However, there are two major differences.
First, neutrino mass is very small, with respect to quark (and charged
lepton) mass. Second, lepton mixing can be large and even maximal, while
quark mixing is small. Both features can be explained by means of the
(type I) seesaw mechanism \cite{ss}.

In fact, for a single fermion generation, the effective neutrino mass
$m_{\nu}$ is given by $m_{\nu} \simeq (m_D/m_R)m_D$, where the Dirac
mass $m_D$ is of the order of the quark (or charged lepton) mass, and
the right-handed Majorana mass $m_R$ is of the order of the unification
or intermediate mass scale. As a result, $m_{\nu}$ is very small with
respect to $m_D$.

For two and three fermion generations, the effective neutrino mass matrix
$M_{\nu}$ is given by the formula
\ini
M_{\nu} \simeq M_D^T M_R^{-1} M_D,
\fin
so that large neutrino mixing can be generated from a nearly diagonal $M_D$:
A strong mass hierarchy or large offdiagonal elements in $M_R$ are required
\cite{smirnov,js1}.
A small contribution to the lepton mixing from the charged lepton mass matrix
$M_e$ is also expected, which could be important to understand the deviation
from maximal mixing \cite{js2}.

The type I seesaw mechanism is based on the introduction, within the minimal
standard model, of three heavy right-handed neutrinos. However, small
neutrino masses can be generated also by the inclusion of a heavy Higgs
triplet \cite{triplet}. In this case the neutrino mass matrix is given by
$M_{\nu}=M_L=Y_L v_L$, where $Y_L$ is a Yukawa matrix and $v_L$ is the
v.e.v. of the triplet, which can be written as $v_L=\gamma v^2/m_T$,
with $v$ the v.e.v. of a standard Higgs doublet, $m_T$ the triplet mass,
and $\gamma$ a coefficient related to the coupling between the doublet
and the triplet. Then, $v_L$ is small with respect to $v$, but
large mixing in $M_{\nu}$ is achieved by hand. Instead, from the type I
seesaw formula (1) we get $M_{\nu} \simeq Y_D^T M_R^{-1} Y_D v^2$, so
that large mixing can be generated from the structure of both matrices
$Y_D$ and $M_R$.

More generally, we can write also a type II seesaw formula by adding to
the usual type I term (1) the triplet (or type II) term, so that
\ini
M_{\nu} \simeq M_D^T M_R^{-1} M_D + M_L.
\fin
This is called type II seesaw mechanism. Large mixing in $M_L$ should
be introduced by hand. Nevertheless, the structure and the scale of $M_L$
could produce an effect on neutrino mixing and even explain the deviation
from maximal mixing.

In the present paper we take as starting point a horizontal $U(2)$
inspired model for fermion mass matrices \cite{ddff}, in order to explore
the type II seesaw mechanism. We aim to study in particular the
deviation from maximal mixing, the contribution of $M_e$ and $M_L$,
and the value of the small mixing angle $\theta_{13}$. Contrary to
Ref.\cite{rode1}, we consider the type I term, and not $M_L$, as the
basic neutrino mass term.

\section{Lepton mixing}

The lepton mixing matrix $U$ is defined by $U=U_e^{\dg} U_{\nu}$, where 
$U_e$ and $U_{\nu}$ diagonalize $M_e$ and $M_{\nu}$, respectively. Since
$U_{23}$ is near maximal, $U_{12}$ is large but not maximal, and $U_{13}$ 
is small, then $U$ has the approximate form
\ini
U \simeq
\left( \begin{array}{ccc}
c & s & \epsilon \\
-\frac{1}{\sqrt2}(s+c \epsilon) & \frac{1}{\sqrt2}(c-s \epsilon) &
\frac{1}{\sqrt2} \\
\frac{1}{\sqrt2}(s-c \epsilon) & -\frac{1}{\sqrt2}(c+s \epsilon) &
\frac{1}{\sqrt2}
\end{array} \right),
\fin
with $s \simeq \frac{1}{\sqrt3}$ and $\epsilon < 0.2$. In fact, assuming 
a standard parametrization, we have
\ini
0.48 < \sin \theta_{12} < 0.62,
\fin
\ini
0.56 < \sin \theta_{23} < 0.84,
\fin
with central values $\sin \theta_{12}=0.55$, $\sin \theta_{23}=0.70$,
respectively, and $\sin \theta_{13} < 0.23$. For a recent clear account
of neutrino phenomenology, see Ref.\cite{alt}.

\section{Mass matrices}

Horizontal symmetries, which relate particles of different generations,
have been often used to explain the hierarchical pattern of mass
matrices \cite{hs}. For instance, the Abelian $U(1)$ symmetry and the
non-Abelian $U(2)$ symmetry. The last one is based on the assumption
that the three generations transform as a doublet plus a singlet:
$\psi_a + \psi_3$. Then the $U(2)$ symmetry is broken down to the
$U(1)$ symmetry and again to nothing, generating specific forms of
mass matrices.

According to the $U(2)$ model described in Ref.\cite{ddff}, we have
the following approximate expression for neutrino mass matrices, 
\ini
M_D \sim
\left( \begin{array}{ccc}
\la^{12} & \la^6 & \la^{10} \\
-\la^6 & \la^4 & \la^4 \\
\la^{10} & \la^4 & 1
\end{array} \right)~m_t,
\fin
\ini
M_R \sim
\left( \begin{array}{ccc}
\la^{12} & \la^{10} & \la^6 \\
\la^{10} & \la^8 & \la^4 \\
\la^6 & \la^4 & 1
\end{array} \right)~m_R,
\fin
where $\la=0.2$ is the Cabibbo parameter, and $m_t \simeq v$.
Moreover, we have
\ini
M_e \sim
\left( \begin{array}{ccc}
\la^{6} & \la^3 & \la^{5} \\
-\la^3 & \la^2 & \la^2 \\
\la^{5} & \la^2 & 1
\end{array} \right)~m_b.
\fin
Matrices (6) and (8) were obtained, on the phenomenological side,
by inserting the quark mass hierarchy into a widely adopted form of
quark mass matrices \cite{qmm}, and, on the theoretical side, by
means of the broken $U(2)$ family symmetry \cite{u2}. Simple
quark-lepton relations, $M_e \sim M_d$ and $M_D \sim M_u$, were also
assumed. Matrix (7) was obtained by inverting the (type I) seesaw
formula, and was found consistent with the broken $U(2)$ horizontal
symmetry.

Then $M_{\nu}^I \simeq M_D^T M_R^{-1} M_D$ may be given by the matrix
\ini
M_{\nu}^I \sim
\left( \begin{array}{ccc}
\la^4 & \la^2 & -\la^2 \\
\la^2 & 1 & 1 \\
-\la^2 & 1 & 1
\end{array} \right)~\frac{m_t^2}{m_R},
\fin
which corresponds to a normal mass hierarchy of neutrinos.
We will take also
\ini
M_{\nu}^{II}=M_L=\frac{m_L}{m_R}M_R.
\fin
The last relation can be motivated by assuming that the structure of both 
matrices $M_R$ and $M_L$ is generated by coupling with the same flavon 
fields. It is just a conjecture: since the Dirac mass matrices have
similar structures, we may think that both the Majorana mass matrices
have one specific structure.

\section{Model exploration}

Since the 2-3 sector of $M_{\nu}^{I}$ could not be exactly democratic,
we consider the following form, as a perturbation of (9),
\ini
M_{\nu}^I \simeq
\left( \begin{array}{ccc}
\la^4 & \la^2 & -\la^2 \\
\la^2 & 1+\frac{\la^n}{2} & 1-\frac{\la^n}{2} \\
-\la^2 & 1-\frac{\la^n}{2}  & 1+\frac{\la^n}{2}
\end{array} \right)~\frac{m_t^2}{m_R},
\fin
which has 2-3 symmetry \cite{lam} and hence near maximal 2-3 mixing,
with $n=1,2,3,4$. Then, for $n=4$ we get the bimaximal mixing, that is
$s=1/\sqrt2$ and $\epsilon=0$.
For $n=3$ we have
$\tan 2 \theta_{12} \simeq 2 \sqrt2 /\la$ or $\sin \theta_{12} \simeq 0.68$.
For $n=2$, 
$\tan 2 \theta_{12} \simeq 2 \sqrt2$ or $\sin \theta_{12} \simeq 0.58$.
For $n=1$, 
$\tan 2 \theta_{12} \simeq 2 \sqrt2 \la$ or $\sin \theta_{12} \simeq 0.25$.

The contribution from $M_{\nu}^{II}$, with respect to $M_{\nu}^{I}$,
will be parametrized by the ratio
\ini
k=\frac{m_L m_R}{v^2}=\gamma ~\frac{m_R}{m_T}
\fin
and leads to a decrease of the mixings. We perform a numerical 
analysis. For useful formulas see the appendix of Ref.\cite{rode2}.

Now we consider the contribution to lepton mixing from $M_e$ \cite{gt,fp}.
This is similar to the Wolfenstein parametrization of the CKM matrix \cite{wolf}.
We easily obtain
\ini
\sin \theta_{12} \simeq
\sin \theta_{12}^{\nu} -\frac{\la}{2},
\fin
\ini
\sin \theta_{23} \simeq \frac{1}{\sqrt2} \left( 1-\la^2 \right),
\fin
\ini
\sin \theta_{13} \simeq -\frac{\la}{\sqrt2}.
\fin
If we include in the element 2-2 of $M_e$ the $-3$ factor by
Georgi and Jarlskog (GJ) \cite{gj}, which reproduces better
$m_e$ and $m_{\mu}$, we get instead
\ini
\sin \theta_{12} \simeq
\sin \theta_{12}^{\nu} +\frac{\la}{6},
\fin
$$
\sin \theta_{23} \simeq \frac{1}{\sqrt2} \left( 1-\la^2 \right),
$$
\ini
\sin \theta_{13} \simeq \frac{\la}{3 \sqrt2}.
\fin
Taking into account the three contributions,
$M_{\nu}^I$, $M_{\nu}^{II}$, and $M_e$, in this order,
and matching with the allowed ranges of lepton mixings,
reported in section II, we get the following results.

Case $n=4$ requires $0 \le k<0.05$, or $0.08<k<0.18$ for the GJ option.

Case $n=3$ requires $0 \le k<0.04$, or $0.06<k<0.16$ for the GJ option.

Case $n=2$ is reliable only for the GJ choice with $0 \le k<0.10$.

Case $n=1$ is not reliable at all.

\newpage

\section{Discussion}

We see that the contribution from $M_{\nu}^{II}$ is necessary
for the cases $n=4$ and $n=3$ with the GJ option.
However, its impact is most important only on the 1-2 mixing, while
it is of the order $10^{-2}$ on the 2-3 mixing and $10^{-3}$ on the
1-3 mixing. Therefore, in our framework, near maximal 2-3 mixing can
be ascribed mainly to $M_{\nu}^I$, and the small but not zero 1-3 mixing
mainly to $M_e$. Instead, three mass matrices contribute to the
1-2 mixing, thus providing naturally a large but not maximal value. 

It has been pointed out \cite{fp} that the inclusion of the $-3$ factor
by GJ is not consistent with the observed quark-lepton complementarity
$\theta_{12} + \theta_{C} \simeq \pi/4$. Based on the previous section,
we argue that the contribution of the triplet seesaw is able to correct
such a disagreement.

We have studied a type I term of the form (11), and a type II term of the 
form (10),(7). For other choices considered in the literature, see for 
instance Ref.\cite{other}.

\end{document}